\documentclass[
reprint,
superscriptaddress,
 showpacs, 
 amsmath, amssymb, 
 aps, pre, 
]{revtex4-1}

\usepackage[dvipdfmx]{graphicx} 
\usepackage{dcolumn}
\usepackage{bm}

\newcommand{\md}{\mathrm{d}}

\begin{document}

\preprint{APS/123-QED}

\title{
Asymmetry in cilia configuration induces hydrodynamic phase locking 
}


\author{Keiji Okumura}
 \email{kokumura@aoni.waseda.jp}
\author{Seiya Nishikawa}
\affiliation{
 Department of Electrical Engineering and Bioscience, 
 Waseda University, 
 Sinjuku-ku, Tokyo 169-8555, Japan}
\author{Toshihiro Omori}
\author{Takuji Ishikawa}
\affiliation{
 Department of Finemechanics, 
 Tohoku University, 
 Sendai, Miyagi 980-8579, Japan}
\author{Atsuko Takamatsu}
\affiliation{
 Department of Electrical Engineering and Bioscience, 
 Waseda University, 
 Sinjuku-ku, Tokyo 169-8555, Japan}

%
%

\date{\today}

\begin{abstract}
To gain insight into the nature of biological synchronization at the microscopic scale, we here investigate the hydrodynamic synchronization between conically rotating objects termed nodal cilia. 
A mechanical model of three rotating cilia is proposed with consideration of variation in their shapes and geometrical arrangement. 
We conduct numerical estimations of both near-field and far-field hydrodynamic interactions, and apply a conventional averaging method for weakly coupled oscillators. 
In the non-identical case, the three cilia showed stable locked-phase differences around $\pm \pi/2$. 
However, such phase locking also occurred with three identical cilia when allocated in a triangle except for the equilateral triangle. 
The effects of inhomogeneity in cilia shapes and geometrical arrangement on such asymmetric interaction is discussed to understand the role of biological variation in synchronization via hydrodynamic interactions. 
\end{abstract}


\maketitle



\section{\label{sec:introduction}Introduction}

Synchronization phenomena are ubiquitously observed in nature from microscopic to macroscopic scales \cite{Pikovsky_Synchronization_2001}. 
Examples of the latter case include the synchronization of animal circadian clocks with the sun \cite{Yamaguchi_Mice_2013}, fireflies simultaneously flashing \cite{Strogatz_Sync_2004}, and frogs calling in anti-phase \cite{Aihara_Modeling_2009}. 
In contrast to the above examples in which the elements interact without materials, synchronization in microscopic organisms involves physical interactions such as hydrodynamic interaction. 
For instance, the cooperative movement of cilia or flagella in fluids is essential for propelling the microorganisms themselves or facilitating the transport of biological particles \cite{Golestanian_Hydrodynamic_2011,Elgeti_Physics_2015}. 
Although it is intuitively acceptable that such synchronization can be achieved more easily in oscillators with identical properties, hydrodynamic interaction induces phenomena that oppose this intuition. 
Indeed, several studies have shown the difficulty of synchronization between two identical cilia coupling through a fluid \cite{Lenz_Collective_2006, Uchida_Hydrodynamic_2012}, whereas non-identical cilia (\textit{e.g.}, cilia with different lengths and/or moving along different orbits) can synchronize more easily \cite{Takamatsu_Hydrodynamic_2013}. 
Here, we address the effect of such inhomogeneity on the establishment of synchronization, considering not only the inhomogeneity of the properties of the oscillators themselves but also the asymmetric geometrical arrangement of the oscillators using a three-cilium system.

\begin{figure}[b]
 \centering
 \includegraphics[width=6.5cm,clip]{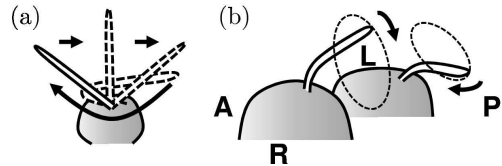}
 \caption{\label{fig:ManyKindsOfCilia} 
 Schematic illustration of two kinds of motile cilia. 
 (a) Beating motion, consisting of an effective stroke and a recovery stroke. 
 (b) Rotating motion in a node; L, R, A, and P indicate the left, right, anterior, and posterior sides of the node cavity in mice, respectively. 
 }
\end{figure}

Cilia or flagella attached to the cell surface exhibit two types of movements in fluids: beating and rotating. 
The beating motion consists of an effective stroke and a recovery stroke, in which the cilia move almost in a plane perpendicular to the cell surface, as shown in Fig.~\ref{fig:ManyKindsOfCilia}(a) \cite{Brumley_Flagellar_2014}. 
In contrast, the rotating motion is conical, as shown in Fig.~\ref{fig:ManyKindsOfCilia}(b). 
However, the majority of theoretical studies on cilia movement have focused on the beating movement with only a few studies focusing on rotating movement. 
Nevertheless, this rotating motion of cilia is biologically important during organ development, specifically for determining the animal body plan, which has been widely demonstrated in different species, including mice \cite{Nonaka_Randomization_1998,Nonaka_Determination_2002}, rabbits \cite{Okada_Mechanism_2005}, rice fish \cite{Okada_Mechanism_2005}, and frogs \cite{Schweickert_Cilia_2007}.

The mouse embryo is the most extensively investigated system for understanding rotating cilia, in which a caved domain, known as a node, is filled with liquid \cite{Nonaka_Randomization_1998}. 
The node consists of a few hundreds of cells, each of which has a cilium conically rotating clockwise when viewed from the top (Fig.~\ref{fig:ManyKindsOfCilia}(b)). 
The rotational axis tilts toward the posterior, resulting in the generation of leftward fluid flow, which has proven to be essential for establishment of left-right (LR) asymmetry at an early developmental stage \cite{Nonaka_Determination_2002}. 
The driving torque was reported to be asymmetric during the rotational period, generating effective- and recovery-like strokes \cite{Takamatsu_Asymmetric_2013}. 
Moreover, Shinohara \textit{et al.} revealed that only a few cilia are sufficient to trigger this LR determination in the mouse embryo \cite{Shinohara_Two_2012}. 
In their experimental setup, the positions of rotating cilia were randomly distributed for both wild type and mutant embryos, including a nearest-neighbor distance near collision. 
They also measured a certain degree of variation in the cilia lengths. 
Takamatsu \textit{et al.} experimentally observed cooperative rotational movement in isolated pairs of cilia in a mutant mouse \cite{Takamatsu_Hydrodynamic_2013}. 
In their study, most of the observed phase lags were detected at a certain phase, when the difference between the natural frequencies of the two cilia tended to be small. 
When the differences were large, phase shifts were observed.

Experimental discoveries in the mouse embryo \cite{Nonaka_Determination_2002} motivated further theoretical explanations on the flow produced by rotating cilia. 
Since the cilia and flagella are sufficiently small and the viscosity around them is sufficiently high, they can be modeled under a low Reynolds number regime. 
In applying the resistive force theory and the slender body theory, Smith {\it et al.} revealed the optimal tilt configuration for generating maximum flow \cite{Smith_Discrete_2007, Smith_Fluid_2008}. 
They estimated the pathline of virtual particles, representing biological signals, under a few or large arrays of identical cilia with the assumption of in-phase and out-of-phase synchronization.

Other theoretical studies addressed the system by approximating the portions of the cilia with far-field interactions via fluid as simple rigid spheres termed Stokeslets \cite{Vilfan_Hydrodynamic_2006,Lenz_Collective_2006,Niedermayer_Synchronization_2008,Brumley_Hydrodynamic_2012,Uchida_Generic_2011,Uchida_Hydrodynamic_2012}. 
These studies showed that identical Stokeslets cannot phase-synchronize under a constant driving force in any trajectory \cite{Lenz_Collective_2006, Uchida_Hydrodynamic_2012}. 
However, synchronization could be achieved by introducing harmonic springs in a radius of a circular trajectory \cite{Niedermayer_Synchronization_2008}, a non-identical orbit between two oscillators \cite{Uchida_Generic_2011}, or a time-varying driving force \cite{Vilfan_Hydrodynamic_2006,Uchida_Generic_2011,Uchida_Hydrodynamic_2012}. 
Golestanian \textit{et al.} pointed out that ``if the system is symmetric under the exchange of the two oscillators, then it cannot synchronize \cite{Golestanian_Hydrodynamic_2011}.'' 
Conversely, asymmetry in the system could be a key factor required for achieving synchronization. 
Indeed, experimental systems have revealed the possibility that cilia are inhomogeneous in their shapes \cite{Shinohara_Two_2012}, and an asymmetrical rotational stroke has been reported \cite{Takamatsu_Asymmetric_2013}. 
Since the geometrical arrangement of node cells in tissues is rather irregular in the mouse \cite{Hashimoto_Planar_2010} relative to that of, for example, the $Drosophila$ eye structure \cite{Zaessinger_Drosophila_2015}, asymmetry in the geometrical configuration itself should be considered in a theoretical framework using a system consisting of more than two oscillators.

These inescapable facts raise the fundamental question as to whether such biological variation contributes to or disrupts hydrodynamic synchronization. 
To address this question, we investigated the cases of two and three rotating cilia interacting through a fluid, taking both the geometrical arrangement and length of the cilia into account. 
We conducted a numerical estimation of the hydrodynamic interaction to explore the role of asymmetry for synchronization of the whole system, considering not only far-field but also near-field interactions \cite{Ishikawa_Hydrodynamic_2006}, followed by adoption of the conventional averaging method for weakly coupled oscillators \cite{Kuramoto_Chemical_1984}.

\section{\label{sec:model}Model}

To consider the shape effect of cilia on hydrodynamic synchronization, Takamatsu \textit{et al.} proposed a mechanical model of two rotating cilia and derived the averaged phase equation \cite{Takamatsu_Asymmetric_2013, Takamatsu_Hydrodynamic_2013}. 
The present study follows this modeling method for three rotating cilia.

\subsection{\label{sec:mechanical}Mechanical model of rotating cilia}

\begin{figure}[tb]
 \centering
 \includegraphics[width=6.0cm,clip]{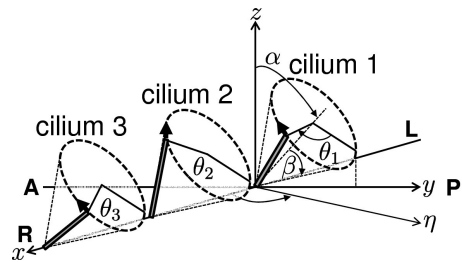}
 \caption{\label{fig:schematic} 
 Coordinate system for the rotational movement of cilia. L, R, A, and P denote the left, right, anterior, and posterior sides in a node cavity, respectively.
 }
\end{figure}

To describe the rotating motion of cilium $i$, let us introduce a rigid cylinder characterized by its length $l_i$ and diameter $r_i$, where $i = 1, 2, 3$ (Figs.~\ref{fig:ManyKindsOfCilia}(b) and~\ref{fig:schematic}). 
The typical unit length is set to $l_0$, which represents a mean cilium length around 
2--3 ${\rm \mu m}$. 
The surface of the node cells is treated as a flat wall. 
Let us suppose that the cylinder rotates on a prescribed orbit of a cone surface, whose center axis tilts to a direction $\eta_i$ with a tilt angle $\alpha_i$ and an open angle $\beta_i$, as shown in Fig.~\ref{fig:schematic}. 
The right-handed Cartesian coordinate system is used so that the positive $x$-axis corresponds to the right side of an embryo. 
The root position of cilium $i$ is denoted as $(x_i, y_i, r_i/2)$ 
\footnote{To avoid a collision between a cilium and the basal wall when $\alpha_i + \beta_i$ equals $\pi/2$, the root of the cilium is raised from the wall by the cylindrical radius as $z_i = r_i/2$. }.
It is assumed that each cilium is driven by a torque.

Let $\theta_i$ be the phase of rotating cilium $i$ defined by the clockwise angle on the conical base, measured from the positive $y$-axis direction. 
We denote the phase vector and the driving torque vector as 
$\bm{\Theta} = (\theta_1, \theta_2, \theta_3)^{\mathrm{T}}$ 
and 
$\bm{T} = (T_1(\theta_1), T_2(\theta_2), T_3(\theta_3))^{\mathrm{T}}$ 
, respectively. 
In the Stokes flow regime, the driving torque acting on cilium $i$ balances the sum of the viscous drag torque generated by cilium $i$ itself and the other cilia as 
\begin{equation}
\bm{T} = K \dot{\bm{\Theta}} 
\label{eq:balance_equation} . 
\end{equation}
Here, 
\begin{equation}
K = (K_{ij})
\label{eq:resistance_matrix}
\end{equation}
is a resistance matrix, representing the hydrodynamic interaction. 
Its component of the drag coefficient $K_{ij}$ from cilium $j$ to $i$ is numerically calculated using the boundary element method (BEM) described below.

The drag coefficient $K_{ij}$ is derived by the flow field generated by the motions of cilia. 
The velocity of the fluid $\bm{v}$ at position $\bm{r}$ can be calculated by superpositioning the velocities generated by every cilium as follows \cite{Pozrikidis_Boundary_1992}: 
\begin{equation}
 \bm{v}(\bm{r}) = - \sum_{i=1}^3 \int_{A_i} G(\bm{r}-\bm{r'}) \bm{q}(\bm{r'}) \textrm{d}A_i, \label{eq:GreenFunction}
\end{equation}
where $\bm{q}$ is a traction force exerted at position $\bm{r'}$, and $A_i$ is the surface of cilium $i$. 
The function $G$ is the half-space Green's function, known as the Blake tensor \cite{Blake_A_1971}, which satisfies the no-slip boundary condition at the flat wall. 
Hereafter, no background flow is considered. 
The surface of the cilia is also under the no-slip boundary condition. 
The velocities of certain positions of the cilium equal those of the surrounding fluid. 
Using this relation and Eq.~(\ref{eq:GreenFunction}), the distribution of traction force $\bm{q}$ exerted on the cilium surface can be estimated.

To estimate $\bm{q}$, we adopt the BEM for the numerical evaluation of Eq.~(\ref{eq:GreenFunction}) \cite{Ishikawa_Hydrodynamic_2006, Takamatsu_Asymmetric_2013, Takamatsu_Hydrodynamic_2013}. 
For instance, let us suppose that cilium 1 rotates while cilia 2 and 3 are fixed at certain phases so that: 
$\dot{\bm{\Theta}} = (1, 0, 0)^{\mathrm{T}}$. 
From Eq.~(\ref{eq:balance_equation}), we have 
\begin{equation*}
(T_1(\theta_1), T_2(\theta_2), T_3(\theta_3))^{\mathrm{T}} 
= 
(K_{11}, K_{21}, K_{31})^{\mathrm{T}}. 
\end{equation*}
The surface of each cilium is discretized into 574 triangular elements. 
The estimation of integration in Eq.~(\ref{eq:GreenFunction}) is conducted on each triangular element using the 4-point Gaussian polynomials. 
The singularity in the integration is fixed numerically \cite{Pozrikidis_Finite_1995}. 
As a consequence, the closed linear algebraic equation is obtained, which can be numerically solved by applying the LU decomposition to obtain the traction force $\bm{q}$. 
The drag coefficient $K_{ij}$ is calculated using the following definition of the drag torque:
\begin{equation}
 T^{(d)}_i = -\bm{e} \cdot \int_{A_i} \bm{q} \times  \bm{r'} \textrm{d}A_i, \label{eq:integral_traction}
\end{equation}
where $\bm{e}$ is the unit vector along the rotational axis of the cone. 
Although the BEM is computationally costly, it allows for precisely estimating the near-field interactions between rotating cilia via fluid, whereas the Stokeslet model can only consider far-field interactions.

To describe the hydrodynamic interaction function explicitly, the resistance matrix is divided into the following forms: 
$K_{ii} = J_i(\theta_i) + \epsilon K'_{ii}(\bm{\Theta})$ 
and 
$K_{ij} = \epsilon K'_{ij}(\bm{\Theta})$, 
where $J_i$ represents the drag coefficient for an isolated single cilium, $K'_{ij}$ represents that for the interaction, and $\epsilon > 0$ is the interaction strength. 
Note that $K'_{ij}$ is a function of not only the phase variables $\theta_i$ and $\theta_j$ but also the third variable $\theta_k$. 
If there is less interaction due to sufficient spacing $\epsilon \approx 0$, the system recovers to the isolated single-cilium system, \textit{i.e.}, $K_{ii} \approx J_i$, which takes a maximum value near the basal wall. 
Since the strength of a hydrodynamic interaction decays according to $1/\textrm{(distance)}^3$ \cite{Blake_A_1971}, $K'_{ij}$ is expected to be considerably small. 
Indeed, the order of $K'_{ij}$ is 100-times smaller than that of $J_i$ in the BEM. 
Thus, the determinant for the resistance matrix is expected to take on a non-zero value 
$|K| \neq 0$, 
so that 
$\dot{\bm{\Theta}} = K^{-1} \bm{T}$. 
Hence, the time evolution for $\theta_i$ in the three-cilium model is written as follows: 
\begin{widetext}
\begin{equation}
\dot{\theta}_i = \frac{1}{|K|}\{T_i J_j J_k 
 + \epsilon [(K'_{jj}T_i - K'_{ij}T_j)J_k + (K'_{kk}T_i - K'_{ik}T_k)J_j] 
 + \epsilon^2 h_i\}, 
\label{eq:ode}
\end{equation}
\end{widetext}
where $h_i = (K'_{jj}K'_{kk} - K'_{jk}K'_{kj})T_i + (K'_{ik}K'_{kj} - K'_{ij}K'_{kk})T_j + (K'_{ij}K'_{jk} - K'_{ik}K'_{jj})T_k$ 
, and the indices are chosen from $\{1, 2, 3\}$ without overlapping. 
For simplicity, the driving torque is assumed to be constant.

\subsection{\label{sec:average}Averaged phase equation}

In this subsection, we analyze the phase synchronization using the system of Eq.~(\ref{eq:ode}). 
In fact, the angular velocity of rotating cilia is not constant over the time evolution of Eq.~(\ref{eq:ode}). 
Even when a single cilium is isolated from others and its driving torque is constant, the cilium moves faster when it is farther from the cell surface owing to the viscous drag torque. 
This can result in the time-varying phase difference between the rotating cilia during a period of a single cycle. 
To apply the conventional phase-averaging method, let us introduce a new definition of phase $\psi_i$ instead of $\theta_i$ for the case of an isolated cilium. 
Without any interactions $\epsilon=0$, 
the phase evolution is written as 
$\dot{\theta}_i = T_i/J_i(\theta_i)$. 
We define the new phase variable $\psi_i$ so that its angular velocity takes on a constant value of natural frequency $\omega_i$ as 
$\dot{\psi_i} := \omega_i$. 
The transformation of the phase variable from $\theta_i$ to $\psi_i$ is expressed as 
\begin{equation}
 \psi_i 
 = \omega_i \int_0^{\theta_i} \frac{1}{\dot{\theta_i'}} \md \theta_i' 
 = \omega_i \int_0^{\theta_i} \frac{J_i(\theta_i')}{T_i} \md \theta_i' . 
 \label{eq:phase_transformation}
\end{equation}
Let 
$\bm{\Psi} = (\psi_1, \psi_2, \psi_3)^{\mathrm{T}}$ 
be the transformed phase vector.

We can now describe the dynamics of hydrodynamically interacting rotating cilia based on the new phase. 
The new drag coefficients corresponding to the phase transformation are denoted as $\hat{J}$ and $\hat{K}'$. 
With interaction 
$\epsilon \neq 0$, 
because 
$J \gg K'_{ii}, K'_{ij}$
and the squared and higher terms are negligible, 
\begin{equation}
 |K| \approx J_1(\theta_1)J_2(\theta_2)J_3(\theta_3). 
 \label{eq:Kapprox}
\end{equation}
By applying the phase transformation (Eq.~(\ref{eq:phase_transformation})) to the time evolution of 
$\theta_i$ (Eq.~(\ref{eq:ode})), 
and using Eq.~(\ref{eq:Kapprox}), we obtain 
\begin{equation*}
\dot{\psi_i} \simeq \omega_i + \epsilon \omega_i [\gamma_{ij}(\bm{\Psi}) + \gamma_{ik}(\bm{\Psi})], 
\end{equation*}
where 
$\gamma_{ij}(\bm{\Psi})$ $:=$ $(\hat{K}'_{jj}(\bm{\Psi})T_i-\hat{K}'_{ij}(\bm{\Psi})T_j)/(\hat{J}_j(\psi_j)T_i)$ 
and 
$\gamma_{ik}(\bm{\Psi})$ $:=$ $(\hat{K}'_{kk}(\bm{\Psi})T_i-\hat{K}'_{ik}(\bm{\Psi})T_k)/(\hat{J}_k(\psi_k)T_i)$. 
Note that this approximation is used only in the analytical framework for synchronization, and is not applied in the numerical calculations presented in Section~\ref{sec:result}.

Let us now consider the phase synchronization of rotating cilia with respect to the transformed phase variable. 
Introducing a phase difference as $\Phi_{ij} := \psi_j - \psi_i$, we obtain 
\begin{eqnarray}
\dot{\Phi}_{ij} &=& \Delta_{ij} + \epsilon \{ \omega_j [\gamma_{ji}(\bm{\Psi}) +\gamma_{jk}(\bm{\Psi})] \nonumber\\
&& - \omega_i [\gamma_{ij}(\bm{\Psi})+\gamma_{ik}(\bm{\Psi})]\}, \nonumber 
\end{eqnarray}
where 
$\Delta_{ij} := \omega_j - \omega_i$ 
is the difference between the natural frequencies. 
It is expected that the time evolution of the phase difference $\Phi_{ij}$ is much slower than that of each phase variable $\psi_i$. 
Thus, by averaging over a period of $\omega_i$ with respect to $\psi_i$ \cite{Kuramoto_Chemical_1984}, 
the following phase equation can be derived: 
\begin{equation}
\dot{\Phi}_{ij} \simeq \Delta_{ij} + \epsilon \Gamma^*_{ij}(\bm{\Phi};\omega_i,\omega_j), 
\label{eq:phase_equation}
\end{equation}
where 
\begin{eqnarray}
\Gamma^*_{ij}(\bm{\Phi};\omega_i,\omega_j) &:=& \omega_j[\Gamma_{ji}(\bm{\Phi})+\Gamma_{jk}(\bm{\Phi})] \nonumber\\
&& - \omega_i[\Gamma_{ij}(\bm{\Phi})+\Gamma_{ik}(\bm{\Phi})], \nonumber\\
\Gamma_{ij}(\bm{\Phi}) &:=& \frac{1}{2\pi} \int_0^{2\pi} \gamma_{ij}(\bm{\Psi}) \md \psi_1. \nonumber
\end{eqnarray}
Here, $\bm{\Phi} = (\Phi_{12}, \Phi_{23})^{\mathrm{T}}$. 
We here investigate the two-dimensional dynamical behavior of the phase difference $\bm{\Phi}$ to study the relationships between phase synchronization and the geometrical arrangement or length in the three-cilium model.

\section{\label{sec:result}Results}

The drag coefficients and the time evolution numerically calculated using the mechanical model described in Section~\ref{sec:mechanical} are presented in Section~\ref{sec:numerical}. 
Subsequently, based on the analytical treatment described in Section~\ref{sec:average}, the effect of the geometrical arrangement or length of cilia on phase locking is presented in Section~\ref{sec:analytical}.

\subsection{\label{sec:numerical}Numerical calculation of the mechanical model}

\begin{figure}[tb]
 \centering
 \includegraphics[width=\linewidth,clip]{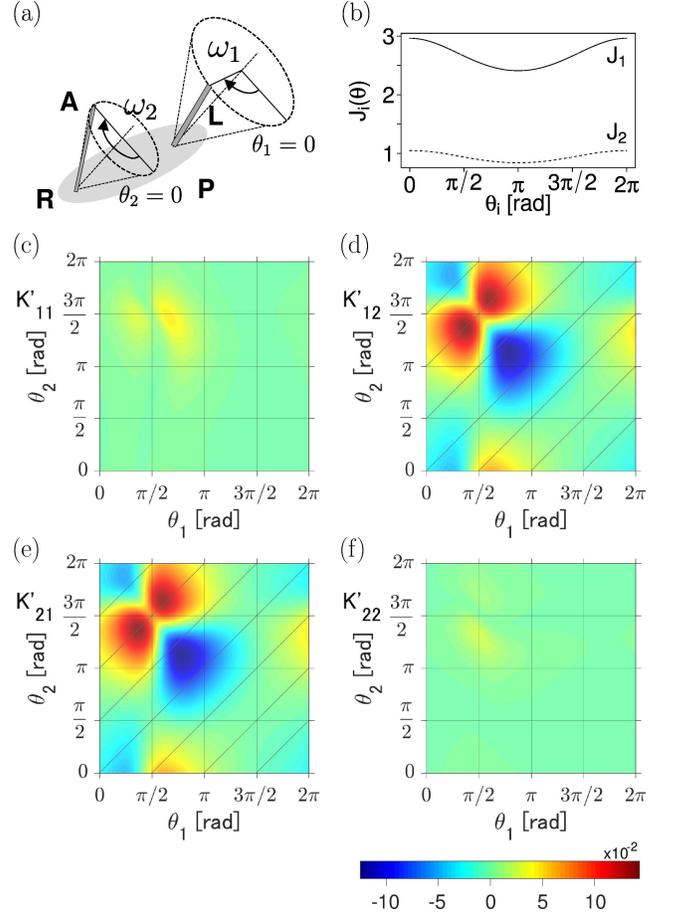}
\caption{\label{fig:C2drag} 
 Numerical estimation of viscous drag coefficients in the two-cilium system. 
 (a) Setup. 
 (b) Drag coefficients $J_i$ for an isolated cilium. 
 (c)--(f) Drag coefficients $K'_{ij}$ as an interaction function from cilium $j$ to $i$. 
 The diagonal lines are included as a visual guide in (d) and (e) (see Section~\ref{sec:discussion}). 
 The parameter values of each cilium are 
 $\alpha_{i}=\pi/6$, $\beta_{i}=\pi/4$, $\eta_{i}=\pi/2$, $r_{i}=0.1l_0$, $l_{1}=1.5l_0$, $l_{2}=l_0=1$, $T_1=0.140$, and $T_2=0.0494$, respectively. 
 The driving torque $T_i$ is set so that the natural frequency $\omega_i$ is 52.4 [rad/s]. 
 The root positions of cilia 1 and 2 are set to the origin and $(x_2, y_2) = (2l_0,0)$, respectively. 
 The drag coefficients are normalized by $\mu \omega l_0^3$, where $\mu$ is the viscosity. 
 }
\end{figure}

\begin{figure*}[p]
 \centering
 \includegraphics[width=\linewidth,clip]{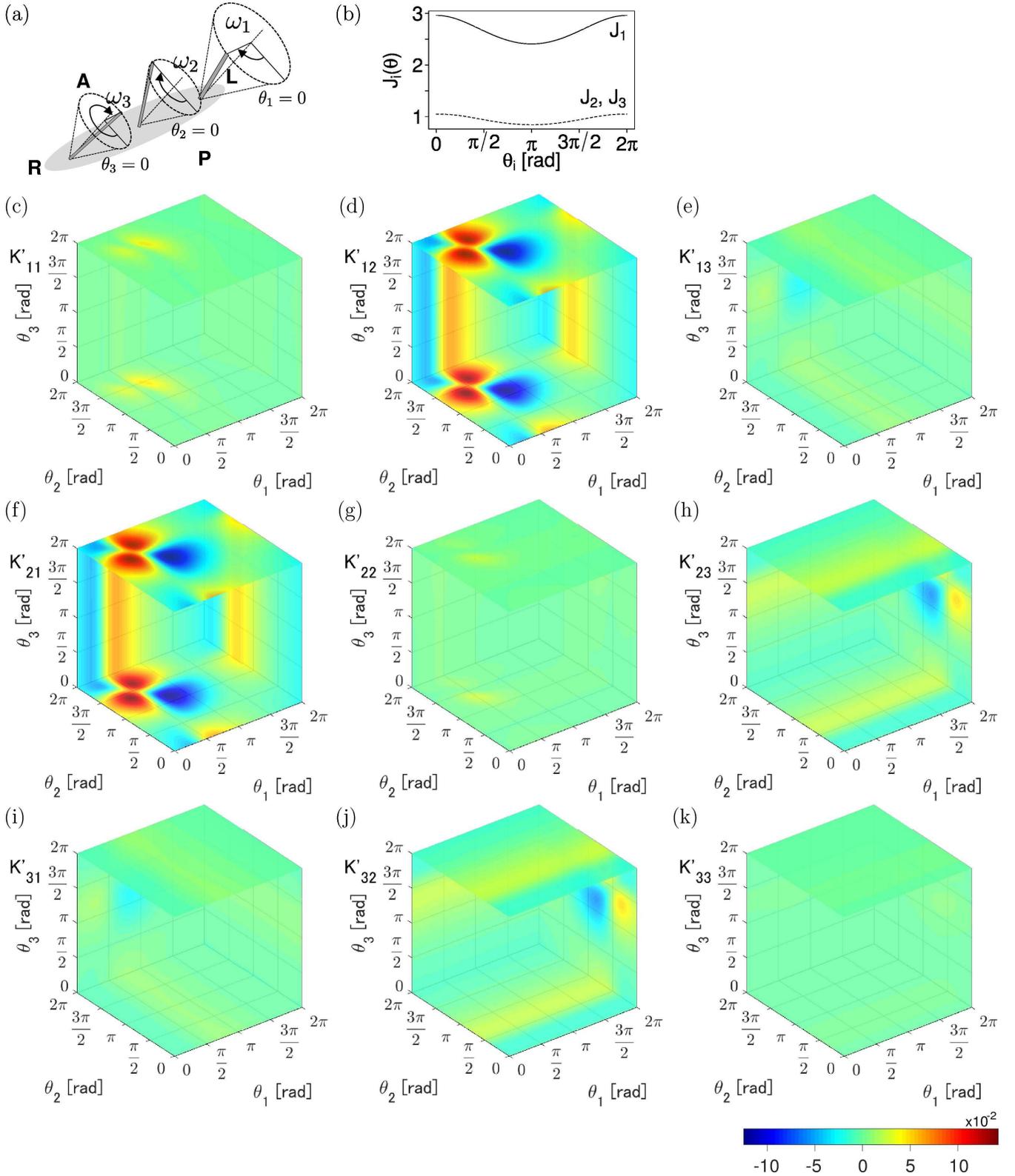}
 \caption{\label{fig:C3drag} 
 Numerical estimation of viscous drag coefficients in the three-cilium system. 
 (a) Setup. 
 (b) Drag coefficients $J_i$. 
 (c)--(k) Drag coefficients $K'_{ij}$. 
 Here, the sections of $K'_{ij}$ at $\theta_1 = 2\pi$, $\theta_2 = 2\pi$, and $\theta_3 = 0, 2\pi$ are displayed. 
 The parameter values of cilium 3 are 
  $\alpha_{3}=\pi/6$, $\beta_{3}=\pi/4$, $\eta_{3}=\pi/2$, $r_{3}=0.1l_0$, $l_{3}=l_0=1$, $T_3=0.0494$, and $(x_3, y_3) = (4l_0,0)$, in addition to those of Fig.~\ref{fig:C2drag}. 
 The drag coefficients are normalized by $\mu \omega l_0^3$, where $\mu$ is the viscosity. 
 }
\end{figure*}

To compare the mechanical model of three rotating cilia with that of two rotating cilia, the drag coefficients $J$ and $K'$ estimated using Eqs.~(\ref{eq:GreenFunction}) and~(\ref{eq:integral_traction}) for both models are presented in Figs.~\ref{fig:C2drag} and~\ref{fig:C3drag}, respectively. 
To analyze the effect of inhomogeneity in cilia shapes, we placed all cilia on the $x$-axis, where the left-most cilium 1 is longer than the others (Figs.~\ref{fig:C2drag}(a) and~\ref{fig:C3drag}(a)). 
The parameter values of $\alpha_i$ and $\beta_i$ are set as typical values according to experimental observation \cite{Shiratori_The_2006}. 
The drag coefficient of isolated cilium $J_1$ is larger than that of $J_{2,3}$ because of the difference in their lengths (Figs.~\ref{fig:C2drag}(b) and~\ref{fig:C3drag}(b)). 
Therefore, the driving torques were set so that all natural frequencies were equal. 
Consequently, $T_1$ is larger than $T_{2,3}$. 
It is obvious that the value of $J_i$ is 100-times larger than that of $K'_{ij}$, resulting in a weakly coupled system (Figs.~\ref{fig:C2drag}(c)--(f) and~\ref{fig:C3drag}(c)--(k)). 
In the two-cilium system, the value of $K'_{ii}$ is approximately zero in almost all domains except for the left upper domain resulting from the existence of the other cilium (Figs.~\ref{fig:C2drag}(c) and~\ref{fig:C2drag}(f)). 
The function $K'_{ij}$ nearly equals $K'_{ji}$ (Figs.~\ref{fig:C2drag}(d) and~\ref{fig:C2drag}(e)). 
These features also hold in the three-cilium system except that the amplitude of function $K'_{ij}$ depends on the distance. 
Note that even though $K'_{13}$ and $K'_{31}$ consider the existence of cilium 2, the effect is minor (Figs.~\ref{fig:C3drag}(e) and~\ref{fig:C3drag}(i)).

\begin{figure}[t]
 \centering
 \includegraphics[width=7.5cm,clip]{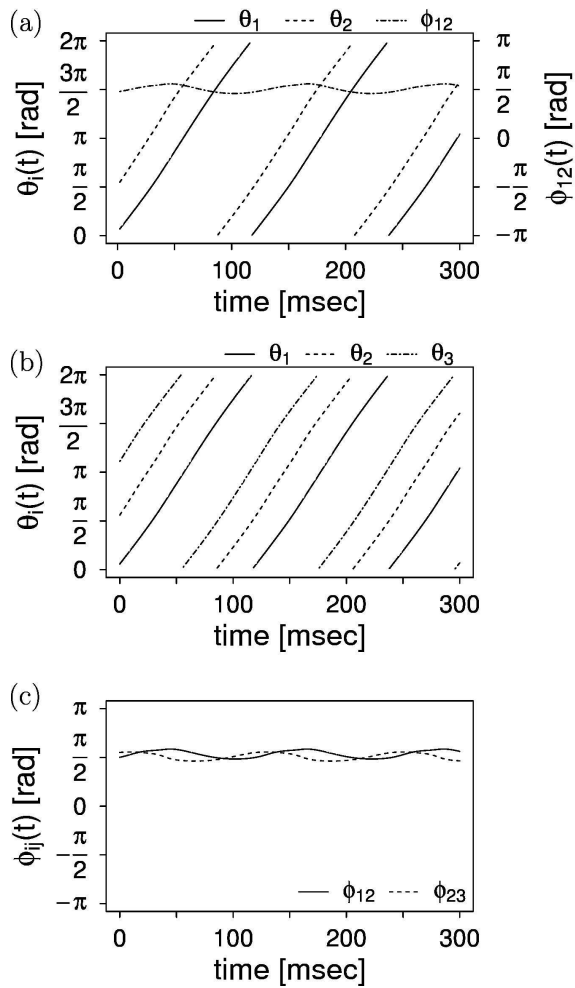}
 \caption{\label{fig:dynamics} 
 Time evolution of the phases of rotating cilia. 
 (a) Evolution of the original phases $\theta_1$ and $\theta_2$ and their difference $\phi_{12} = \theta_2 - \theta_1$ in the two-cilium system. 
 (b) Evolution of the original phases $\theta_1$, $\theta_2$, and $\theta_3$ in the three-cilium system. 
 (c) Evolution of the phase differences $\phi_{12} = \theta_2 - \theta_1$ and $\phi_{23} = \theta_3 - \theta_2$. 
 The cilium parameter values are the same as those defined in Figs.~\ref{fig:C2drag} and ~\ref{fig:C3drag} in the two- and three-cilium systems, respectively. 
}
\end{figure}

Figure~\ref{fig:dynamics} shows the time evolution of each phase $\theta_i$ and the phase difference $\phi_{ij}$ $:=$ $\theta_j - \theta_i$ in the two- and three-cilium models based on Eq.~(\ref{eq:ode}). 
As shown in Fig.~\ref{fig:dynamics}(a), the shorter cilium ($\theta_2$) rotates in advance of the longer cilium ($\theta_1$) around $\pi/2$ in the two-cilium system. 
In the three-cilium system, the phase differences also tend to be around $\pi/2$ (Figs.~\ref{fig:dynamics}(b) and~\ref{fig:dynamics}(c)). 
These results were independent of the initial phase values. 
In both the two- and three-cilium model, the convergence time for phase locking from a random initial phase was typically within a few tens of a beating cycle, corresponding to a few seconds in an experiment.

\subsection{\label{sec:analytical}Effect of the geometrical arrangement of cilia on phase locking analyzed by the averaging method}

\begin{figure}[tb]
 \centering
 \includegraphics[width=6.5cm,clip]{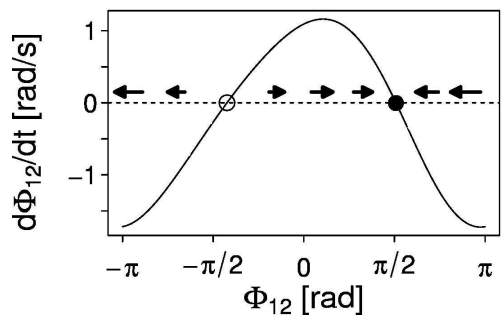}
 \caption{\label{fig:C2PRC} 
 Phase portrait of the average phase difference $\Phi_{12}$ in the $(\Phi_{12}, \dot{\Phi}_{12})$ plane. 
 Stable and unstable fixed points are indicated by filled and open circles, respectively. 
 The parameter values of the cilia are the same as those given in Fig.~\ref{fig:C2drag}. 
}
\end{figure}

Figure~\ref{fig:C2PRC} represents the phase portrait of the two-cilium system using the averaged phase equation (Eq.~(\ref{eq:phase_equation})), suggesting that the phase-locked state occurs around $\pi/2$. 
This state can be achieved if the natural frequency difference $\Delta_{ij}$ satisfies the following condition: 
\begin{equation}
\min \Gamma^*_{ij} < -\Delta_{ij} / \epsilon < \max \Gamma^*_{ij}. 
\label{eq:deltaomega}
\end{equation}
Therefore, the maximum and minimum values of $\Gamma^*_{ij}$ can be indices for the range realizing the phase-locked state with respect to $\Delta_{ij}$. 
We also evaluated the eigenvalue $\lambda$ at the fixed points, which corresponds to the slope in this case. 
At the stable fixed point, the absolute inverse value of $\lambda$ relates to the convergence time for phase locking. 
The value of $\lambda$ was estimated as 
-1.68 $\rm{s^{-1}}$, 
suggesting that the phase-locked state can be achieved within a few tens of the rotating cycle period. 

\begin{figure*}[tb]
 \centering
 \includegraphics[width=\linewidth,clip]{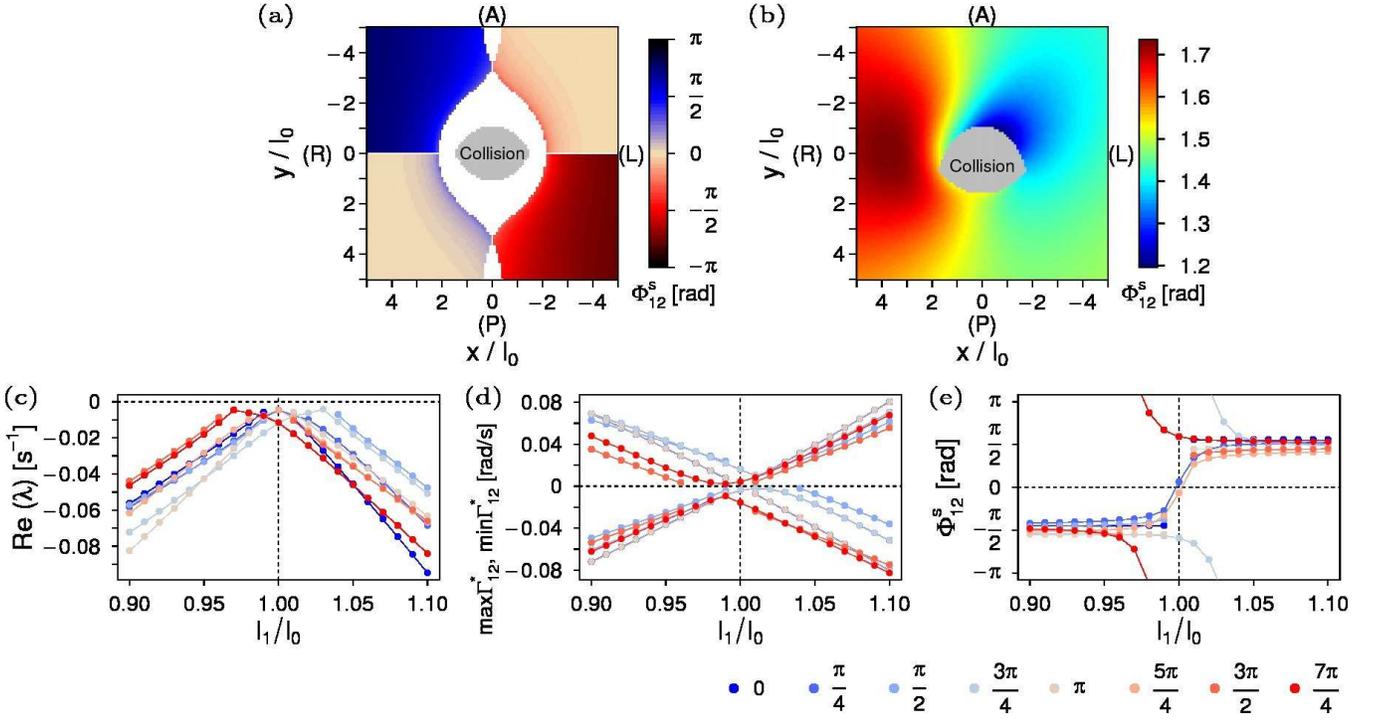}
 \caption{\label{fig:PhaseDifferenceGap} 
 Dependence of phase-locked states on arrangement and shapes in the two-cilium model. 
 (a),(b) Spatial dependence in the identical and non-identical (different cilia lengths) settings. 
 Cilium 1 was placed at the origin and cilium 2 as placed at an arbitrary position. 
 The white area indicates the asynchronous states. 
 (c)--(e) Phase-locked properties with a change in length. 
 Cilium 2 was placed at the points with radius $3l_0$ and angle $n\pi/4$ $(n=0,...,7)$. 
 The cilium parameter values are the same as those described in Fig.~\ref{fig:C2drag} except for $l_{1}=l_0$ and $T_1=0.0494$ in (a). 
 }
\end{figure*}

To comprehend the effects of inhomogeneity in cilia shapes and geometrical arrangement on phase locking in the two-cilium model, we conducted a systematic investigation as shown in Fig.~\ref{fig:PhaseDifferenceGap}. 
Even in the system consisting of two identical cilia, the geometrical arrangement clearly contributed to phase locking, except for the arrangement on the $LR$-axis and the $AP$-axis (Fig.~\ref{fig:PhaseDifferenceGap}(a)). 
When the length of cilium 1 was set to $l_1 = 1.5l_0$ (non-identical case), the locked-phase differences changed to certain values distinguished from the in-phase or anti-phase (Fig.~\ref{fig:PhaseDifferenceGap}(b)). 
As shown in Figs.~\ref{fig:PhaseDifferenceGap}(c)--(e), even a slight change in length (less than $5\%$) results in phase-locked states at around $\pm \pi/2$ with decreasing $\operatorname{Re}(\lambda)$ and an increasing range of $\Gamma^{*}_{12}$. 
This finding was independent of the geometrical arrangement. 
Overall, these results indicate that phase locking is easily induced by the difference in cilia lengths, compared with the geometrical arrangement in the two-cilium system.

\begin{figure*}[t]
 \centering
 \includegraphics[width=\linewidth,clip]{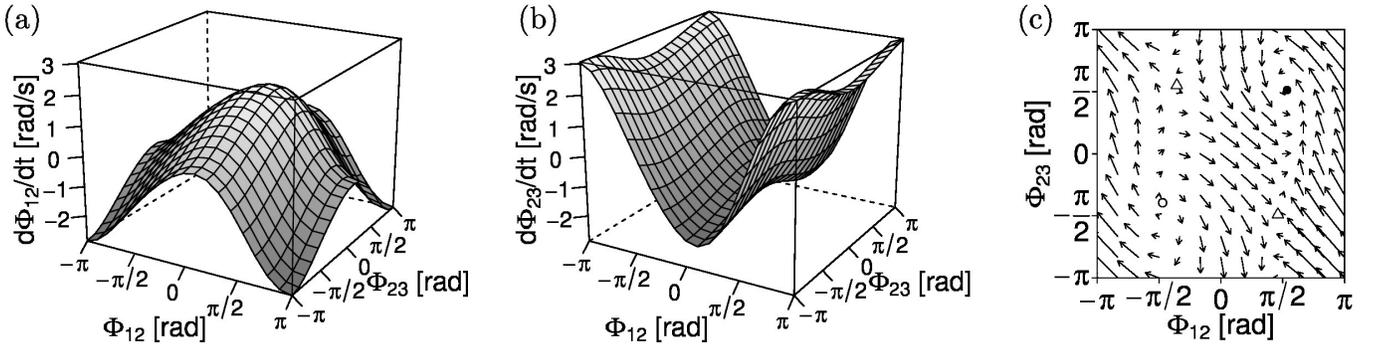}
 \caption{\label{fig:C3PRC} 
 Phase portrait of the average phase difference in the three-cilium system. 
 (a),(b) Phase portraits of the average phase differences $\Phi_{12}$ and $\Phi_{23}$ in the $(\Phi_{12}, \Phi_{23}, \dot{\Phi}_{12})$ and $(\Phi_{12}, \Phi_{23}, \dot{\Phi}_{23})$ spaces, respectively. 
 (c) Vector field in the phase space $(\Phi_{12}, \Phi_{23})$, summarizing the results of phase space analyses in (a) and (b). 
 The filled and open circles represent stable and unstable foci, respectively. 
 Open triangles denote saddlepoints. 
 The cilium parameter values are the same as those given in Fig.~\ref{fig:C3drag}. 
}
\end{figure*}

Figure~\ref{fig:C3PRC} shows the result of phase plane analysis in the three-cilium system. 
Figures~\ref{fig:C3PRC}(a) and~\ref{fig:C3PRC}(b) denote the phase portraits of $\Phi_{12}$ and $\Phi_{23}$, which provide information on their time evolution. 
By combining these two portraits, the vector field in phase space $(\Phi_{12}, \Phi_{23})$ and fixed points are obtained (Fig.~\ref{fig:C3PRC}(c)). 
Here, only the complete phase-locked state, in which all cilia are phase-locked, is considered. 
Then, conducting linear stability analysis, we found that the stable fixed point is around $(\Phi_{12}, \Phi_{23}) = (\pi/2, \pi/2)$, which is consistent with the result of the numerical simulation for the phase difference $\phi_i$ calculated using the original phase equation (Eq.~(\ref{eq:ode})), as presented in Fig.~\ref{fig:dynamics}(c). 
The largest real part of the eigenvalues at the stable fixed point is evaluated as 
-0.84 $\rm{s^{-1}}$. 
As seen in Figs.~\ref{fig:C2PRC} and~\ref{fig:C3PRC}(c), the range of $\dot{\Phi}_{12}$ in the three-cilium system is wider than that in the two-cilium system. 
This means that phase locking can be encouraged even if the differences between the natural frequencies of the former system are relatively large.

\begin{figure*}[!t]
 \centering
 \includegraphics[width=\linewidth,clip]{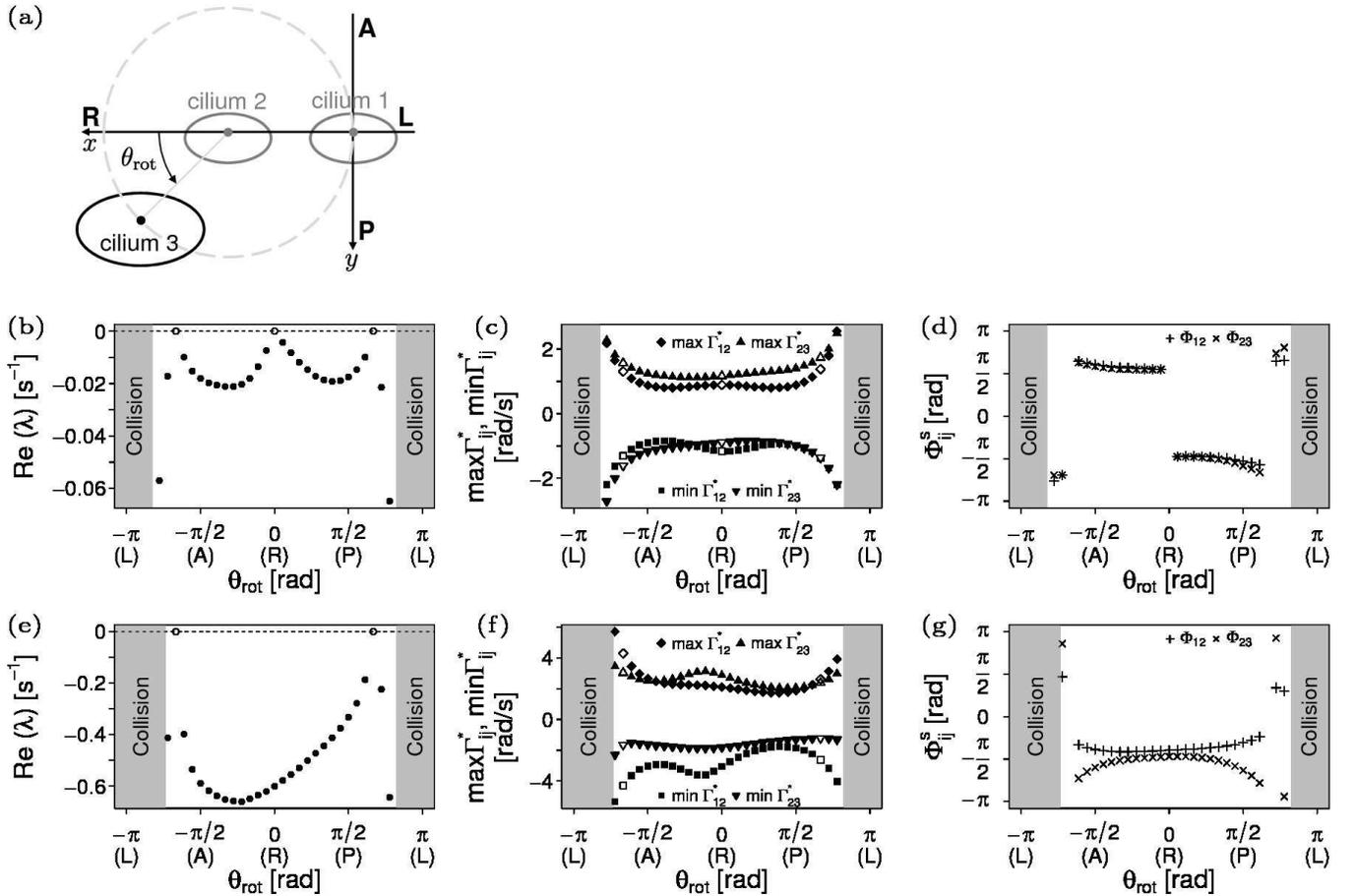}
 \caption{\label{fig:Arrangement} 
 Effects of geometrical arrangement on phase locking. 
 (a) Spatial arrangement of the three cilia. 
 (b)--(d) Results for the identical settings. 
 (e)--(g) Results for the non-identical (different cilia lengths) settings. 
 (b),(e) The largest real part of the eigenvalues.
 (c),(f) Range of the natural frequency difference. 
 Open and filled circles indicate asynchronous and phase-locked states, respectively. 
 (d),(g) Average locked-phase difference. 
 The parameter values of the cilia are the same as those in Fig.~\ref{fig:C3drag} except for 
 $l_{1}=l_0$ and $T_1=0.0494$ 
 in the identical situation, and 
 $l_{1}=l_0$, $T_1=0.0494$, $l_{3}=1.5l_0$, and $T_3=0.140$ 
 in the non-identical situation. 
 In both cases, the position of cilium 3 was $(x_3, y_3) = (2l_0\cos\theta_{\rm rot}+2l_0, 2l_0\sin\theta_{\rm rot})$. 
}
\end{figure*}

Finally, the effect of geometrical arrangement in the three-cilium system was investigated. 
Two of the cilia, 1 and 2, were placed at positions on the LR-axis ($x$-axis), where cilium 2 was located at the center of a circle represented by a broken line in Fig.~\ref{fig:Arrangement}(a). 
The third cilium, 3, was placed at a position on the circle, whose coordinate was defined as 
$(x_3, y_3)$ $=$ $(2l_0\cos\theta_{\rm rot}+2l_0, 2l_0\sin\theta_{\rm rot})$ 
using the parameter $\theta_{\rm rot}$. 
The domain of $\theta_{\rm rot}$ is restricted from a collision. 
Figure~\ref{fig:Arrangement}(b)--(g) shows the effect of the geometrical arrangement on phase locking, represented using the following characteristic values: $\operatorname{Re}(\lambda)$, $\max \Gamma^{*}_{ij}$, $\min \Gamma^{*}_{ij}$, and the locked-phase difference $\Phi_{ij}^{\rm s}$. 
When the identical cilia are allocated at symmetric positions forming a line ($\theta_{\rm rot} = 0$) or an equilateral triangle ($\theta_{\rm rot} = \pm \pi/3$), phase locking cannot occur (Fig.~\ref{fig:Arrangement}(b)). 
Precisely, in these situations, $\operatorname{Re}(\lambda)$ are infinitesimally zero $\mathcal O (10^{-8})$ 
\footnote{It cannot be identified whether these are positive or negative owing to numerical error. }
, suggesting that the fixed point is a center. 
The time evolution of the original phase $\theta_{i}$ corresponding to these cases did not show phase locking. 
By contrast, when the identical cilia are placed at the vertices of isosceles-triangles, phase locking is achieved. 
In addition, when the third cilium comes close to the position of cilium 1, $\operatorname{Re}(\lambda)$ goes to negative infinity, suggesting fast phase locking owing to the effect of the near-field interaction. 
Note that in some isosceles triangles ($\theta_{\rm rot}=$ $\pm 11\pi/18$, $\pm 13\pi/18$, $\pm 14\pi/18$), all combinations of the partial two-cilium system did not exhibit phase locking, as shown in Fig.~\ref{fig:PhaseDifferenceGap}(a). 
It should also be noted that the phase-locked states appear around $\pm \pi/2$ in the phase difference, as seen in Fig.~\ref{fig:Arrangement}(d). 
When the third cilium is not identical to the others (\textit{e.g.}, it has a longer length), phase locking is realized, similar to the two-cilium system \cite{Takamatsu_Hydrodynamic_2013}, except for the equilateral-triangular arrangement (Fig.~\ref{fig:Arrangement}(e)). 
Compared with the case of identical cilia, the phase locking is easily established as confirmed with the values of $\operatorname{Re}(\lambda)$, which typically decreased from -0.02 (Fig.~\ref{fig:Arrangement}(b)) to -0.6 (Fig.~\ref{fig:Arrangement}(e)), and the range of $\Gamma^{*}_{ij}$, which typically doubled, as shown in Fig.~\ref{fig:Arrangement}(f) compared with Fig.~\ref{fig:Arrangement}(c). 
The locked-phase differences were detected at around $- \pi/2$ (Fig.~\ref{fig:Arrangement}(g)).

\section{\label{sec:discussion}Discussion}

\begin{figure}[!tb]
 \centering
 \includegraphics[width=8.0cm,clip]{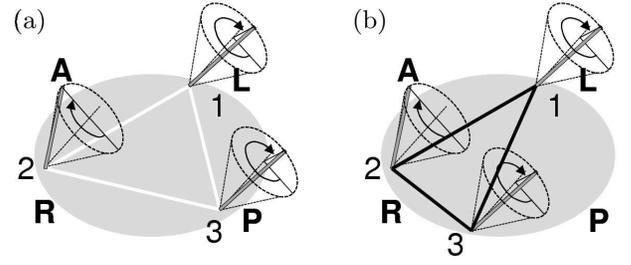}
 \caption{\label{fig:ArrangementIllustration} 
 Schematic illustration of the arrangement of three identical cilia for symmetric and asymmetric interactions. 
 (a) Equilateral-triangular arrangement. 
 (b) Isosceles-triangular arrangement. 
 }
\end{figure}

We proposed a mechanical model of three rotating cilia to investigate the effects of geometrical arrangement and inhomogeneity in cilia shapes, which was inspired by experimentally observed biological variation. 
We found that inhomogeneity in cilia shapes causes phase locking in the three-cilium system (Figs.~\ref{fig:dynamics} and~\ref{fig:C3PRC}) as well as in the two-cilium system \cite{Takamatsu_Hydrodynamic_2013}. 
Interestingly, even the system consisting of three `identical cilia' is able to be phase-locked when the cilia are placed at the vertices of a triangle except for the equilateral triangle (Fig.~\ref{fig:ArrangementIllustration}). 
In this section, we discuss the scenario of phase locking, focusing on asymmetry in interactions: inhomogeneity in cilia shapes and asymmetry in geometrical arrangement.

First, we consider the inhomogeneity in cilia shapes in the two-cilium system. 
We modify Eq.~(\ref{eq:ode}) into that of the two-cilium model. 
Then, the evolution of phase difference can be written as follows 
\footnote{Note that because the effect of transformation from $\theta_i$ to $\psi_i$ is slight, a qualitative understanding is possible by using the original phases $\theta_{1,2}$ for simplicity. }: 
\begin{eqnarray}
  \dot{\theta}_{2} - \dot{\theta}_{1} &=& \frac{1}{|K|} [(K_{11}T_2 - K_{22}T_1) - (K_{21}T_1 - K_{12}T_2)] \nonumber\\
  &\simeq& \Delta_{12} - \frac{\epsilon}{J_1 J_2} (K'_{21}T_1 - K'_{12}T_2), 
 \label{eq:discuss_C2}
\end{eqnarray}
where the approximations $|K| \approx J_1 J_2$ (see also Eq.~(\ref{eq:Kapprox}))  and $K'_{ii} \approx 0$ are used. 
To realize the phase-locked state, the rotating speed is assumed to satisfy $\Delta_{12} = \omega_2 - \omega_1 = 0$, for simplicity. 
When the length of cilium $i$ is longer, $J_i$ becomes larger (Fig.~\ref{fig:C2drag}). 
To compensate for this larger drag torque, the driving torques need to be $T_i > T_j$. 
In contrast, as shown in Figs.~\ref{fig:C2drag}(d) and~\ref{fig:C2drag}(e), $K'_{12} \approx K'_{21}$. 
Consequently, $K'_{21}T_1 - K'_{12}T_2$ in Eq.~(\ref{eq:discuss_C2}) becomes non-zero. 
By averaging this term along the line denoted by $\theta_2 = \theta_1 + \phi_{12}$ (represented with the diagonal lines in Figs.~\ref{fig:C2drag}(d) and~\ref{fig:C2drag}(e)) over the period of $\theta_1$, the cosine-type function, similar to that of Fig.~\ref{fig:C2PRC}, is obtained. 
Here, $\phi_{12}$ is an arbitrary phase difference. 
When this cosine-type function gets across zero, the phase-locked state is achieved. 
In the case of two identical cilia (\textit{e.g.}, placed on the $LR$-axis), the drag coefficient $K'_{ij}$ is symmetric such as 
\begin{equation*}
 K'_{12}(\theta_1,\theta_2) = K'_{21}(-\theta_2,-\theta_1), 
\end{equation*}
which means that $K'_{21}$ is the reflection of $K'_{12}$ in the $\theta_2 = 2\pi -\theta_1$ axis 
\footnote{In addition, $K'_{12}(\theta_1,\theta_2) \approx K'_{21}(\theta_1,\theta_2)$ also holds. Precisely, $K'_{12}(\theta_1,\theta_2) \neq K'_{21}(\theta_1,\theta_2)$, since the drag torque from cilium 1 to 2 does not equal that from cilium 2 to 1 under the model considering the cilium length. 
}. 
Therefore, the sum of $K'_{ij}T_j$ for a single rotating cycle is the same for both cilia 1 and 2, resulting in no phase locking \cite{Golestanian_Hydrodynamic_2011}.

Second, we discuss the effect of asymmetry in the geometrical arrangement for the three-cilium system. 
Similar to the case of the two-cilium system, by using Eq.~(\ref{eq:ode}), the time evolution of phase differences is described as 
\begin{eqnarray}
  \dot{\theta}_{2} - \dot{\theta}_{1} &\simeq& 
   \Delta_{12} - \frac{\epsilon}{J_1 J_2} (K'_{21}T_1 - K'_{12}T_2) \nonumber\\
   &&- \epsilon (\frac{K'_{23}T_3}{J_2 J_3} - \frac{K'_{13}T_3}{J_1 J_3}), \label{eq:discuss_C3_1} \\
  \dot{\theta}_{3} - \dot{\theta}_{2} &\simeq& 
   \Delta_{23} - \frac{\epsilon}{J_2 J_3} (K'_{32}T_2 - K'_{23}T_3) \nonumber\\
   &&- \epsilon (\frac{K'_{31}T_1}{J_1 J_3} - \frac{K'_{21}T_1}{J_1 J_2}). 
 \label{eq:discuss_C3_2}
\end{eqnarray}
Let us suppose that these cilia shapes are identical. 
Thus, $T_1$ $=$ $T_2$ $=$ $T_3$ is assumed to realize the same natural frequencies. 
The second terms in the right side of Eqs.~(\ref{eq:discuss_C3_1}) and~(\ref{eq:discuss_C3_2}), which correspond to the interactions between cilia 1 and 2 and between cilia 2 and 3, nearly equal zero as in the two-cilium system. 
When the three cilia are allocated at the vertices of an equilateral triangle (Fig.~\ref{fig:ArrangementIllustration}(a)), the drag coefficient $K'_{ij}$ is symmetric such as 
\begin{eqnarray*}
 &K'_{12}(\bm{\Theta})=K'_{21}(\bm{\Theta_{{\rm A}}}), 
 &K'_{13}(\bm{\Theta})=K'_{23}(\bm{\Theta_{{\rm A}}}), \\
 &K'_{23}(\bm{\Theta})=K'_{32}(\bm{\Theta_{{\rm B}}}), 
 &K'_{21}(\bm{\Theta})=K'_{31}(\bm{\Theta_{{\rm B}}}), \\
 &K'_{31}(\bm{\Theta})=K'_{32}(\bm{\Theta_{{\rm A}}}), &
\end{eqnarray*}
where
\begin{eqnarray*}
 \bm{\Theta_{{\rm A}}} &=& (-\theta_2, -\theta_1, -\theta_3)^{\mathrm{T}}, \\
 \bm{\Theta_{{\rm B}}} &=& (\frac{2\pi}{3}-\theta_1, \frac{2\pi}{3}-\theta_3, \frac{2\pi}{3}-\theta_2)^{\mathrm{T}}. 
\end{eqnarray*}
Therefore, they cannot be phase-locked. 
By contrast, in the isosceles-triangular arrangement (Fig.~\ref{fig:ArrangementIllustration}(b)), this symmetry is broken, owing to the different distances between cilia 1 and 2 and between cilia 2 and 3. 
This results in phase locking even if all cilia are identical. 
Taken together, these findings indicate that asymmetry in the interaction caused by inhomogeneity in cilia shapes and/or geometrical arrangement induces phase locking.

Here, we focused on the scenario of specific parameter choice with the constraint of the same natural frequencies in the three rotating cilia, so as to best reveal the effects of inhomogeneity in cilia shapes and geometrical arrangement on phase locking. 
Although we also investigated some other parameter values to ensure the generality of our findings, it would not be practical to investigate various values of $\alpha_i$, $\beta_i$, and $\eta_i$ in precise, owing to the high computational cost of estimating the hydrodynamic interaction. 
However, we expect that our results would hold under $\omega_i =$~(const.), since the phase-locked properties in the three-cilium system are partly analogous to those in the two-cilium system, as shown in Figs.~\ref{fig:C2PRC} and \ref{fig:C3PRC}. 
In the two-cilium model, although the difference in $l_i$ greatly affected the occurrence of phase locking, the qualitative aspects were almost identical for the differences in $\alpha_i$ and $\beta_i$ \cite{Takamatsu_Hydrodynamic_2013}. 
In addition, the phase-locked states depended on different values of $\eta_i$. 
We also investigated some non-zero values of $\Delta_{ij}$, which satisfied Eq.~(\ref{eq:deltaomega}). 
At the stable fixed points in the two-cilium model, a non-zero $\Delta_{ij}$ value increased $\operatorname{Re}(\lambda)$ and shifted the locked-phase difference $\Phi_{ij}^{\rm s}$. 
The same effect was observed in the three-cilium system. 
Interestingly, by contrast, with an equilateral-triangular arrangement of the three identical cilia, non-zero $\Delta_{ij}$ led to phase locking. 
Thus, the differences in $\alpha_i$, $\beta_i$, $\eta_i$, and $\omega_i$ in the three cilia would also be expected as a source of inhomogeneity, which could stabilize the phase locking.

Next we compare our model with the Stokeslet model. 
In our study, shapes of cilia were explicitly introduced with the near-field interaction. 
Figure~\ref{fig:PhaseDifferenceGap}(a) is similar to the result reported by Vilfan and J{\"u}licher \cite{Vilfan_Hydrodynamic_2006}, who investigated hydrodynamic phase synchronization in the two-Stokeslet model, neglecting the near-field interaction. 
Our model showed that in the case of identical cilia, $\operatorname{Re}(\lambda)$ and the range of $\Gamma^{*}_{12}$ were extremely small, as exemplified in Figs.~\ref{fig:PhaseDifferenceGap}(c) and~\ref{fig:PhaseDifferenceGap}(d) when $l_1/l_0 = 1$. 
Therefore, experimental observation of hydrodynamic phase synchronization might be difficult for two identical cilia. 
By contrast, it was revealed that inhomogeneity in cilia shapes led to more ready phase locking.

Orbit flexibility is another factor required to establish hydrodynamic phase synchronization in rotating cilia. 
This issue has long been discussed in many studies. 
If the degree of symmetry in a hydrodynamic interaction is high, phase synchronization is impossible \cite{Lenz_Collective_2006}. 
Introducing flexibility as a degree of freedom results in irreversible dynamical behavior, showing complete in-phase synchronization \cite{Niedermayer_Synchronization_2008,Brumley_Flagellar_2014}. 
Although no flexibility was considered in our three-cilium model, introducing flexibility to the two-cilium model was shown to cause a shift of the phase-locked state \cite{Takamatsu_Hydrodynamic_2013}. 
Since the observed locked-phase difference was neither in-phase nor anti-phase in the experiment, the tangential component of the hydrodynamic interaction would govern the system of the nodal cilia in the mouse embryo rather than the radial component as flexibility.

The physiological effects of phase locking in the nodal cilia on the LR determination have not been elucidated to date. 
Shinohara \textit{et. al.} showed that to trigger the left-right asymmetric gene expression, as few as only two cilia are required at the very early developmental stage \cite{Shinohara_Two_2012}. 
In the two-cilium model, Takamatsu \textit{et al.} reported that phase locking increased the maximum phase speed of the shorter cilium \cite{Takamatsu_Hydrodynamic_2013}. 
The same effect was observed in the three-cilium model. 
Another effect of phase locking is to contribute to the sustainment of directional flow. 
In the mouse node, it is considered that immotile cilia located in the boundary of the ventral node cavity functions to sense the mechanical stimuli by the flow. 
Therefore, it might be important to receive a sustained hydrodynamic influence caused by the phase lagged effective strokes of rotating cilia.

To conclude, we have shown that a high degree of symmetry in cilia configuration could prevent hydrodynamically induced phase locking in nodal cilia. 
In an actual biological environment, a certain range of variation in the configuration would be crucial to cause hydrodynamic phase locking.

\begin{acknowledgments}
The authors would like to thank Prof. Hamada (RIKEN) and Prof. Shinohara (Tokyo University of Agriculture and Technology) for valuable discussions. 
This work was supported by CREST, Japan Science and Technology Agency. 
\end{acknowledgments}



\begin{thebibliography}{34}%
\makeatletter
\providecommand \@ifxundefined [1]{%
 \@ifx{#1\undefined}
}%
\providecommand \@ifnum [1]{%
 \ifnum #1\expandafter \@firstoftwo
 \else \expandafter \@secondoftwo
 \fi
}%
\providecommand \@ifx [1]{%
 \ifx #1\expandafter \@firstoftwo
 \else \expandafter \@secondoftwo
 \fi
}%
\providecommand \natexlab [1]{#1}%
\providecommand \enquote  [1]{``#1''}%
\providecommand \bibnamefont  [1]{#1}%
\providecommand \bibfnamefont [1]{#1}%
\providecommand \citenamefont [1]{#1}%
\providecommand \href@noop [0]{\@secondoftwo}%
\providecommand \href [0]{\begingroup \@sanitize@url \@href}%
\providecommand \@href[1]{\@@startlink{#1}\@@href}%
\providecommand \@@href[1]{\endgroup#1\@@endlink}%
\providecommand \@sanitize@url [0]{\catcode `\\12\catcode `\$12\catcode
  `\&12\catcode `\#12\catcode `\^12\catcode `\_12\catcode `\%12\relax}%
\providecommand \@@startlink[1]{}%
\providecommand \@@endlink[0]{}%
\providecommand \url  [0]{\begingroup\@sanitize@url \@url }%
\providecommand \@url [1]{\endgroup\@href {#1}{\urlprefix }}%
\providecommand \urlprefix  [0]{URL }%
\providecommand \Eprint [0]{\href }%
\providecommand \doibase [0]{http://dx.doi.org/}%
\providecommand \selectlanguage [0]{\@gobble}%
\providecommand \bibinfo  [0]{\@secondoftwo}%
\providecommand \bibfield  [0]{\@secondoftwo}%
\providecommand \translation [1]{[#1]}%
\providecommand \BibitemOpen [0]{}%
\providecommand \bibitemStop [0]{}%
\providecommand \bibitemNoStop [0]{.\EOS\space}%
\providecommand \EOS [0]{\spacefactor3000\relax}%
\providecommand \BibitemShut  [1]{\csname bibitem#1\endcsname}%
\let\auto@bib@innerbib\@empty
\bibitem [{\citenamefont {Pikovsky}\ \emph {et~al.}(2001)\citenamefont
  {Pikovsky}, \citenamefont {Rosenblum},\ and\ \citenamefont
  {Kurths}}]{Pikovsky_Synchronization_2001}%
  \BibitemOpen
  \bibfield  {author} {\bibinfo {author} {\bibfnamefont {A.}~\bibnamefont
  {Pikovsky}}, \bibinfo {author} {\bibfnamefont {M.}~\bibnamefont {Rosenblum}},
  \ and\ \bibinfo {author} {\bibfnamefont {J.}~\bibnamefont {Kurths}},\ }\href
  {\doibase 10.1017/cbo9780511755743} {\emph {\bibinfo {title}
  {Synchronization: A universal concept in nonlinear sciences}}}\ (\bibinfo
  {publisher} {Cambridge University Press},\ \bibinfo {year}
  {2001})\BibitemShut {NoStop}%
\bibitem [{\citenamefont {Yamaguchi}\ \emph {et~al.}(2013)\citenamefont
  {Yamaguchi}, \citenamefont {Suzuki}, \citenamefont {Mizoro}, \citenamefont
  {Kori}, \citenamefont {Okada}, \citenamefont {Chen}, \citenamefont {Fustin},
  \citenamefont {Yamazaki}, \citenamefont {Mizuguchi}, \citenamefont {Zhang},
  \citenamefont {Dong}, \citenamefont {Tsujimoto}, \citenamefont {Okuno},
  \citenamefont {Doi},\ and\ \citenamefont {Okamura}}]{Yamaguchi_Mice_2013}%
  \BibitemOpen
  \bibfield  {author} {\bibinfo {author} {\bibfnamefont {Y.}~\bibnamefont
  {Yamaguchi}}, \bibinfo {author} {\bibfnamefont {T.}~\bibnamefont {Suzuki}},
  \bibinfo {author} {\bibfnamefont {Y.}~\bibnamefont {Mizoro}}, \bibinfo
  {author} {\bibfnamefont {H.}~\bibnamefont {Kori}}, \bibinfo {author}
  {\bibfnamefont {K.}~\bibnamefont {Okada}}, \bibinfo {author} {\bibfnamefont
  {Y.}~\bibnamefont {Chen}}, \bibinfo {author} {\bibfnamefont {J.}~\bibnamefont
  {Fustin}}, \bibinfo {author} {\bibfnamefont {F.}~\bibnamefont {Yamazaki}},
  \bibinfo {author} {\bibfnamefont {N.}~\bibnamefont {Mizuguchi}}, \bibinfo
  {author} {\bibfnamefont {J.}~\bibnamefont {Zhang}}, \bibinfo {author}
  {\bibfnamefont {X.}~\bibnamefont {Dong}}, \bibinfo {author} {\bibfnamefont
  {G.}~\bibnamefont {Tsujimoto}}, \bibinfo {author} {\bibfnamefont
  {Y.}~\bibnamefont {Okuno}}, \bibinfo {author} {\bibfnamefont
  {M.}~\bibnamefont {Doi}}, \ and\ \bibinfo {author} {\bibfnamefont
  {H.}~\bibnamefont {Okamura}},\ }\href {\doibase 10.1126/science.1238599}
  {\bibfield  {journal} {\bibinfo  {journal} {Science}\ }\textbf {\bibinfo
  {volume} {342}},\ \bibinfo {pages} {85} (\bibinfo {year} {2013})}\BibitemShut
  {NoStop}%
\bibitem [{\citenamefont {Strogatz}(2004)}]{Strogatz_Sync_2004}%
  \BibitemOpen
  \bibfield  {author} {\bibinfo {author} {\bibfnamefont {S.}~\bibnamefont
  {Strogatz}},\ }\href@noop {} {\emph {\bibinfo {title} {Sync: The Emerging
  Science of Spontaneous Order}}}\ (\bibinfo  {publisher} {Penguin},\ \bibinfo
  {year} {2004})\BibitemShut {NoStop}%
\bibitem [{\citenamefont {Aihara}(2009)}]{Aihara_Modeling_2009}%
  \BibitemOpen
  \bibfield  {author} {\bibinfo {author} {\bibfnamefont {I.}~\bibnamefont
  {Aihara}},\ }\href {\doibase 10.1103/PhysRevE.80.011918} {\bibfield
  {journal} {\bibinfo  {journal} {Phys. Rev. E}\ }\textbf {\bibinfo {volume}
  {80}},\ \bibinfo {pages} {011918} (\bibinfo {year} {2009})}\BibitemShut
  {NoStop}%
\bibitem [{\citenamefont {Golestanian}\ \emph {et~al.}(2011)\citenamefont
  {Golestanian}, \citenamefont {Yeomans},\ and\ \citenamefont
  {Uchida}}]{Golestanian_Hydrodynamic_2011}%
  \BibitemOpen
  \bibfield  {author} {\bibinfo {author} {\bibfnamefont {R.}~\bibnamefont
  {Golestanian}}, \bibinfo {author} {\bibfnamefont {J.}~\bibnamefont
  {Yeomans}}, \ and\ \bibinfo {author} {\bibfnamefont {N.}~\bibnamefont
  {Uchida}},\ }\href {\doibase 10.1039/C0SM01121E} {\bibfield  {journal}
  {\bibinfo  {journal} {Soft Matter}\ }\textbf {\bibinfo {volume} {7}},\
  \bibinfo {pages} {3074} (\bibinfo {year} {2011})}\BibitemShut {NoStop}%
\bibitem [{\citenamefont {Elgeti}\ \emph {et~al.}(2015)\citenamefont {Elgeti},
  \citenamefont {Winkler},\ and\ \citenamefont
  {Gompper}}]{Elgeti_Physics_2015}%
  \BibitemOpen
  \bibfield  {author} {\bibinfo {author} {\bibfnamefont {J.}~\bibnamefont
  {Elgeti}}, \bibinfo {author} {\bibfnamefont {R.}~\bibnamefont {Winkler}}, \
  and\ \bibinfo {author} {\bibfnamefont {G.}~\bibnamefont {Gompper}},\ }\href
  {\doibase 10.1088/0034-4885/78/5/056601} {\bibfield  {journal} {\bibinfo
  {journal} {Rep. Prog. Phys.}\ }\textbf {\bibinfo {volume} {78}},\ \bibinfo
  {pages} {056601} (\bibinfo {year} {2015})}\BibitemShut {NoStop}%
\bibitem [{\citenamefont {Lenz}\ and\ \citenamefont
  {Ryskin}(2006)}]{Lenz_Collective_2006}%
  \BibitemOpen
  \bibfield  {author} {\bibinfo {author} {\bibfnamefont {P.}~\bibnamefont
  {Lenz}}\ and\ \bibinfo {author} {\bibfnamefont {A.}~\bibnamefont {Ryskin}},\
  }\href {\doibase 10.1088/1478-3975/3/4/006} {\bibfield  {journal} {\bibinfo
  {journal} {Phys. Biol.}\ }\textbf {\bibinfo {volume} {3}},\ \bibinfo {pages}
  {285} (\bibinfo {year} {2006})}\BibitemShut {NoStop}%
\bibitem [{\citenamefont {Uchida}\ and\ \citenamefont
  {Golestanian}(2012)}]{Uchida_Hydrodynamic_2012}%
  \BibitemOpen
  \bibfield  {author} {\bibinfo {author} {\bibfnamefont {N.}~\bibnamefont
  {Uchida}}\ and\ \bibinfo {author} {\bibfnamefont {R.}~\bibnamefont
  {Golestanian}},\ }\href {\doibase 10.1140/epje/i2012-12135-5} {\bibfield
  {journal} {\bibinfo  {journal} {Euro. Phys. J. E}\ }\textbf {\bibinfo
  {volume} {35}},\ \bibinfo {pages} {9813} (\bibinfo {year}
  {2012})}\BibitemShut {NoStop}%
\bibitem [{\citenamefont {Takamatsu}\ \emph
  {et~al.}(2013{\natexlab{a}})\citenamefont {Takamatsu}, \citenamefont
  {Shinohara}, \citenamefont {Ishikawa},\ and\ \citenamefont
  {Hamada}}]{Takamatsu_Hydrodynamic_2013}%
  \BibitemOpen
  \bibfield  {author} {\bibinfo {author} {\bibfnamefont {A.}~\bibnamefont
  {Takamatsu}}, \bibinfo {author} {\bibfnamefont {K.}~\bibnamefont
  {Shinohara}}, \bibinfo {author} {\bibfnamefont {T.}~\bibnamefont {Ishikawa}},
  \ and\ \bibinfo {author} {\bibfnamefont {H.}~\bibnamefont {Hamada}},\ }\href
  {\doibase 10.1103/PhysRevLett.110.248107} {\bibfield  {journal} {\bibinfo
  {journal} {Phys. Rev. Lett.}\ }\textbf {\bibinfo {volume} {110}},\ \bibinfo
  {pages} {248107} (\bibinfo {year} {2013}{\natexlab{a}})}\BibitemShut
  {NoStop}%
\bibitem [{\citenamefont {Brumley}\ \emph {et~al.}(2014)\citenamefont
  {Brumley}, \citenamefont {Wan}, \citenamefont {Polin},\ and\ \citenamefont
  {Goldstein}}]{Brumley_Flagellar_2014}%
  \BibitemOpen
  \bibfield  {author} {\bibinfo {author} {\bibfnamefont {D.}~\bibnamefont
  {Brumley}}, \bibinfo {author} {\bibfnamefont {K.}~\bibnamefont {Wan}},
  \bibinfo {author} {\bibfnamefont {M.}~\bibnamefont {Polin}}, \ and\ \bibinfo
  {author} {\bibfnamefont {R.}~\bibnamefont {Goldstein}},\ }\href {\doibase
  10.7554/elife.02750} {\bibfield  {journal} {\bibinfo  {journal} {Elife}\
  }\textbf {\bibinfo {volume} {3}},\ \bibinfo {pages} {e02750} (\bibinfo {year}
  {2014})}\BibitemShut {NoStop}%
\bibitem [{\citenamefont {Nonaka}\ \emph {et~al.}(1998)\citenamefont {Nonaka},
  \citenamefont {Tanaka}, \citenamefont {Okada}, \citenamefont {Takeda},
  \citenamefont {Harada}, \citenamefont {Kanai}, \citenamefont {Kido},\ and\
  \citenamefont {Hirokawa}}]{Nonaka_Randomization_1998}%
  \BibitemOpen
  \bibfield  {author} {\bibinfo {author} {\bibfnamefont {S.}~\bibnamefont
  {Nonaka}}, \bibinfo {author} {\bibfnamefont {Y.}~\bibnamefont {Tanaka}},
  \bibinfo {author} {\bibfnamefont {Y.}~\bibnamefont {Okada}}, \bibinfo
  {author} {\bibfnamefont {S.}~\bibnamefont {Takeda}}, \bibinfo {author}
  {\bibfnamefont {A.}~\bibnamefont {Harada}}, \bibinfo {author} {\bibfnamefont
  {Y.}~\bibnamefont {Kanai}}, \bibinfo {author} {\bibfnamefont
  {M.}~\bibnamefont {Kido}}, \ and\ \bibinfo {author} {\bibfnamefont
  {N.}~\bibnamefont {Hirokawa}},\ }\href {\doibase
  10.1016/S0092-8674(00)81705-5} {\bibfield  {journal} {\bibinfo  {journal}
  {Cell}\ }\textbf {\bibinfo {volume} {95}},\ \bibinfo {pages} {829} (\bibinfo
  {year} {1998})}\BibitemShut {NoStop}%
\bibitem [{\citenamefont {Nonaka}\ \emph {et~al.}(2002)\citenamefont {Nonaka},
  \citenamefont {Shiratori}, \citenamefont {Saijoh},\ and\ \citenamefont
  {Hamada}}]{Nonaka_Determination_2002}%
  \BibitemOpen
  \bibfield  {author} {\bibinfo {author} {\bibfnamefont {S.}~\bibnamefont
  {Nonaka}}, \bibinfo {author} {\bibfnamefont {H.}~\bibnamefont {Shiratori}},
  \bibinfo {author} {\bibfnamefont {Y.}~\bibnamefont {Saijoh}}, \ and\ \bibinfo
  {author} {\bibfnamefont {H.}~\bibnamefont {Hamada}},\ }\href {\doibase
  10.1038/nature00849} {\bibfield  {journal} {\bibinfo  {journal} {Nature}\
  }\textbf {\bibinfo {volume} {418}},\ \bibinfo {pages} {96} (\bibinfo {year}
  {2002})}\BibitemShut {NoStop}%
\bibitem [{\citenamefont {Okada}\ \emph {et~al.}(2005)\citenamefont {Okada},
  \citenamefont {Takeda}, \citenamefont {Tanaka}, \citenamefont {Belmonte},\
  and\ \citenamefont {Hirokawa}}]{Okada_Mechanism_2005}%
  \BibitemOpen
  \bibfield  {author} {\bibinfo {author} {\bibfnamefont {Y.}~\bibnamefont
  {Okada}}, \bibinfo {author} {\bibfnamefont {S.}~\bibnamefont {Takeda}},
  \bibinfo {author} {\bibfnamefont {Y.}~\bibnamefont {Tanaka}}, \bibinfo
  {author} {\bibfnamefont {J.}~\bibnamefont {Belmonte}}, \ and\ \bibinfo
  {author} {\bibfnamefont {N.}~\bibnamefont {Hirokawa}},\ }\href {\doibase
  10.1016/j.cell.2005.04.008} {\bibfield  {journal} {\bibinfo  {journal}
  {Cell}\ }\textbf {\bibinfo {volume} {121}},\ \bibinfo {pages} {633} (\bibinfo
  {year} {2005})}\BibitemShut {NoStop}%
\bibitem [{\citenamefont {Schweickert}\ \emph {et~al.}(2007)\citenamefont
  {Schweickert}, \citenamefont {Weber}, \citenamefont {Beyer}, \citenamefont
  {Vick}, \citenamefont {Bogusch}, \citenamefont {Feistel},\ and\ \citenamefont
  {Blum}}]{Schweickert_Cilia_2007}%
  \BibitemOpen
  \bibfield  {author} {\bibinfo {author} {\bibfnamefont {A.}~\bibnamefont
  {Schweickert}}, \bibinfo {author} {\bibfnamefont {T.}~\bibnamefont {Weber}},
  \bibinfo {author} {\bibfnamefont {T.}~\bibnamefont {Beyer}}, \bibinfo
  {author} {\bibfnamefont {P.}~\bibnamefont {Vick}}, \bibinfo {author}
  {\bibfnamefont {S.}~\bibnamefont {Bogusch}}, \bibinfo {author} {\bibfnamefont
  {K.}~\bibnamefont {Feistel}}, \ and\ \bibinfo {author} {\bibfnamefont
  {M.}~\bibnamefont {Blum}},\ }\href {\doibase 10.1016/j.cub.2006.10.067}
  {\bibfield  {journal} {\bibinfo  {journal} {Curr. Biol.}\ }\textbf {\bibinfo
  {volume} {17}},\ \bibinfo {pages} {60} (\bibinfo {year} {2007})}\BibitemShut
  {NoStop}%
\bibitem [{\citenamefont {Takamatsu}\ \emph
  {et~al.}(2013{\natexlab{b}})\citenamefont {Takamatsu}, \citenamefont
  {Ishikawa}, \citenamefont {Shinohara},\ and\ \citenamefont
  {Hamada}}]{Takamatsu_Asymmetric_2013}%
  \BibitemOpen
  \bibfield  {author} {\bibinfo {author} {\bibfnamefont {A.}~\bibnamefont
  {Takamatsu}}, \bibinfo {author} {\bibfnamefont {T.}~\bibnamefont {Ishikawa}},
  \bibinfo {author} {\bibfnamefont {K.}~\bibnamefont {Shinohara}}, \ and\
  \bibinfo {author} {\bibfnamefont {H.}~\bibnamefont {Hamada}},\ }\href
  {\doibase 10.1103/PhysRevE.87.050701} {\bibfield  {journal} {\bibinfo
  {journal} {Phys. Rev. E}\ }\textbf {\bibinfo {volume} {87}},\ \bibinfo
  {pages} {050701} (\bibinfo {year} {2013}{\natexlab{b}})}\BibitemShut
  {NoStop}%
\bibitem [{\citenamefont {Shinohara}\ \emph {et~al.}(2012)\citenamefont
  {Shinohara}, \citenamefont {Kawasumi}, \citenamefont {Takamatsu},
  \citenamefont {Yoshiba}, \citenamefont {Botilde}, \citenamefont {Motoyama},
  \citenamefont {Reith}, \citenamefont {Durand}, \citenamefont {Shiratori},\
  and\ \citenamefont {Hamada}}]{Shinohara_Two_2012}%
  \BibitemOpen
  \bibfield  {author} {\bibinfo {author} {\bibfnamefont {K.}~\bibnamefont
  {Shinohara}}, \bibinfo {author} {\bibfnamefont {A.}~\bibnamefont {Kawasumi}},
  \bibinfo {author} {\bibfnamefont {A.}~\bibnamefont {Takamatsu}}, \bibinfo
  {author} {\bibfnamefont {S.}~\bibnamefont {Yoshiba}}, \bibinfo {author}
  {\bibfnamefont {Y.}~\bibnamefont {Botilde}}, \bibinfo {author} {\bibfnamefont
  {N.}~\bibnamefont {Motoyama}}, \bibinfo {author} {\bibfnamefont
  {W.}~\bibnamefont {Reith}}, \bibinfo {author} {\bibfnamefont
  {B.}~\bibnamefont {Durand}}, \bibinfo {author} {\bibfnamefont
  {H.}~\bibnamefont {Shiratori}}, \ and\ \bibinfo {author} {\bibfnamefont
  {H.}~\bibnamefont {Hamada}},\ }\href {\doibase 10.1038/ncomms1624} {\bibfield
   {journal} {\bibinfo  {journal} {Nat. Commun.}\ }\textbf {\bibinfo {volume}
  {3}},\ \bibinfo {pages} {622} (\bibinfo {year} {2012})}\BibitemShut {NoStop}%
\bibitem [{\citenamefont {Smith}\ \emph {et~al.}(2007)\citenamefont {Smith},
  \citenamefont {Gaffney},\ and\ \citenamefont {Blake}}]{Smith_Discrete_2007}%
  \BibitemOpen
  \bibfield  {author} {\bibinfo {author} {\bibfnamefont {D.}~\bibnamefont
  {Smith}}, \bibinfo {author} {\bibfnamefont {E.}~\bibnamefont {Gaffney}}, \
  and\ \bibinfo {author} {\bibfnamefont {J.}~\bibnamefont {Blake}},\ }\href
  {\doibase 10.1007/s11538-006-9172-y} {\bibfield  {journal} {\bibinfo
  {journal} {B. Math. Biol.}\ }\textbf {\bibinfo {volume} {69}},\ \bibinfo
  {pages} {1477} (\bibinfo {year} {2007})}\BibitemShut {NoStop}%
\bibitem [{\citenamefont {Smith}\ \emph {et~al.}(2008)\citenamefont {Smith},
  \citenamefont {Blake},\ and\ \citenamefont {Gaffney}}]{Smith_Fluid_2008}%
  \BibitemOpen
  \bibfield  {author} {\bibinfo {author} {\bibfnamefont {D.}~\bibnamefont
  {Smith}}, \bibinfo {author} {\bibfnamefont {J.}~\bibnamefont {Blake}}, \ and\
  \bibinfo {author} {\bibfnamefont {E.}~\bibnamefont {Gaffney}},\ }\href
  {\doibase 10.1098/rsif.2007.1306} {\bibfield  {journal} {\bibinfo  {journal}
  {J. R. Soc. Interface}\ }\textbf {\bibinfo {volume} {5}},\ \bibinfo {pages}
  {567} (\bibinfo {year} {2008})}\BibitemShut {NoStop}%
\bibitem [{\citenamefont {Vilfan}\ and\ \citenamefont
  {J{\"u}licher}(2006)}]{Vilfan_Hydrodynamic_2006}%
  \BibitemOpen
  \bibfield  {author} {\bibinfo {author} {\bibfnamefont {A.}~\bibnamefont
  {Vilfan}}\ and\ \bibinfo {author} {\bibfnamefont {F.}~\bibnamefont
  {J{\"u}licher}},\ }\href {\doibase 10.1103/PhysRevLett.96.058102} {\bibfield
  {journal} {\bibinfo  {journal} {Phys. Rev. Lett.}\ }\textbf {\bibinfo
  {volume} {96}},\ \bibinfo {pages} {058102} (\bibinfo {year}
  {2006})}\BibitemShut {NoStop}%
\bibitem [{\citenamefont {Niedermayer}\ \emph {et~al.}(2008)\citenamefont
  {Niedermayer}, \citenamefont {Eckhardt},\ and\ \citenamefont
  {Lenz}}]{Niedermayer_Synchronization_2008}%
  \BibitemOpen
  \bibfield  {author} {\bibinfo {author} {\bibfnamefont {T.}~\bibnamefont
  {Niedermayer}}, \bibinfo {author} {\bibfnamefont {B.}~\bibnamefont
  {Eckhardt}}, \ and\ \bibinfo {author} {\bibfnamefont {P.}~\bibnamefont
  {Lenz}},\ }\href {\doibase 10.1063/1.2956984} {\bibfield  {journal} {\bibinfo
   {journal} {Chaos}\ }\textbf {\bibinfo {volume} {18}},\ \bibinfo {pages}
  {037128} (\bibinfo {year} {2008})}\BibitemShut {NoStop}%
\bibitem [{\citenamefont {Brumley}\ \emph {et~al.}(2012)\citenamefont
  {Brumley}, \citenamefont {Polin}, \citenamefont {Pedley},\ and\ \citenamefont
  {Goldstein}}]{Brumley_Hydrodynamic_2012}%
  \BibitemOpen
  \bibfield  {author} {\bibinfo {author} {\bibfnamefont {D.~R.}\ \bibnamefont
  {Brumley}}, \bibinfo {author} {\bibfnamefont {M.}~\bibnamefont {Polin}},
  \bibinfo {author} {\bibfnamefont {T.~J.}\ \bibnamefont {Pedley}}, \ and\
  \bibinfo {author} {\bibfnamefont {R.~E.}\ \bibnamefont {Goldstein}},\ }\href
  {\doibase 10.1103/PhysRevLett.109.268102} {\bibfield  {journal} {\bibinfo
  {journal} {Phys. Rev. Lett.}\ }\textbf {\bibinfo {volume} {109}},\ \bibinfo
  {pages} {268102} (\bibinfo {year} {2012})}\BibitemShut {NoStop}%
\bibitem [{\citenamefont {Uchida}\ and\ \citenamefont
  {Golestanian}(2011)}]{Uchida_Generic_2011}%
  \BibitemOpen
  \bibfield  {author} {\bibinfo {author} {\bibfnamefont {N.}~\bibnamefont
  {Uchida}}\ and\ \bibinfo {author} {\bibfnamefont {R.}~\bibnamefont
  {Golestanian}},\ }\href {\doibase 10.1103/PhysRevLett.106.058104} {\bibfield
  {journal} {\bibinfo  {journal} {Phys. Rev. Lett.}\ }\textbf {\bibinfo
  {volume} {106}},\ \bibinfo {pages} {058104} (\bibinfo {year}
  {2011})}\BibitemShut {NoStop}%
\bibitem [{\citenamefont {Hashimoto}\ \emph {et~al.}(2010)\citenamefont
  {Hashimoto}, \citenamefont {Shinohara}, \citenamefont {Wang}, \citenamefont
  {Ikeuchi}, \citenamefont {Yoshiba}, \citenamefont {Meno}, \citenamefont
  {Nonaka}, \citenamefont {Takada}, \citenamefont {Hatta}, \citenamefont
  {Anthony},\ and\ \citenamefont {Hamada}}]{Hashimoto_Planar_2010}%
  \BibitemOpen
  \bibfield  {author} {\bibinfo {author} {\bibfnamefont {M.}~\bibnamefont
  {Hashimoto}}, \bibinfo {author} {\bibfnamefont {K.}~\bibnamefont
  {Shinohara}}, \bibinfo {author} {\bibfnamefont {J.}~\bibnamefont {Wang}},
  \bibinfo {author} {\bibfnamefont {S.}~\bibnamefont {Ikeuchi}}, \bibinfo
  {author} {\bibfnamefont {S.}~\bibnamefont {Yoshiba}}, \bibinfo {author}
  {\bibfnamefont {C.}~\bibnamefont {Meno}}, \bibinfo {author} {\bibfnamefont
  {S.}~\bibnamefont {Nonaka}}, \bibinfo {author} {\bibfnamefont
  {S.}~\bibnamefont {Takada}}, \bibinfo {author} {\bibfnamefont
  {K.}~\bibnamefont {Hatta}}, \bibinfo {author} {\bibfnamefont
  {W.}~\bibnamefont {Anthony}}, \ and\ \bibinfo {author} {\bibfnamefont
  {H.}~\bibnamefont {Hamada}},\ }\href {\doibase 10.1038/ncb2020} {\bibfield
  {journal} {\bibinfo  {journal} {Nat. Cell Biol.}\ }\textbf {\bibinfo {volume}
  {12}},\ \bibinfo {pages} {170} (\bibinfo {year} {2010})}\BibitemShut
  {NoStop}%
\bibitem [{\citenamefont {Zaessinger}\ \emph {et~al.}(2015)\citenamefont
  {Zaessinger}, \citenamefont {Zhou}, \citenamefont {Bray}, \citenamefont
  {Tapon},\ and\ \citenamefont {Djiane}}]{Zaessinger_Drosophila_2015}%
  \BibitemOpen
  \bibfield  {author} {\bibinfo {author} {\bibfnamefont {S.}~\bibnamefont
  {Zaessinger}}, \bibinfo {author} {\bibfnamefont {Y.}~\bibnamefont {Zhou}},
  \bibinfo {author} {\bibfnamefont {S.}~\bibnamefont {Bray}}, \bibinfo {author}
  {\bibfnamefont {N.}~\bibnamefont {Tapon}}, \ and\ \bibinfo {author}
  {\bibfnamefont {A.}~\bibnamefont {Djiane}},\ }\href {\doibase
  10.1242/dev.116277} {\bibfield  {journal} {\bibinfo  {journal} {Development}\
  }\textbf {\bibinfo {volume} {142}},\ \bibinfo {pages} {1102} (\bibinfo {year}
  {2015})}\BibitemShut {NoStop}%
\bibitem [{\citenamefont {Ishikawa}\ \emph {et~al.}(2006)\citenamefont
  {Ishikawa}, \citenamefont {Simmonds},\ and\ \citenamefont
  {Pedley}}]{Ishikawa_Hydrodynamic_2006}%
  \BibitemOpen
  \bibfield  {author} {\bibinfo {author} {\bibfnamefont {T.}~\bibnamefont
  {Ishikawa}}, \bibinfo {author} {\bibfnamefont {M.}~\bibnamefont {Simmonds}},
  \ and\ \bibinfo {author} {\bibfnamefont {T.}~\bibnamefont {Pedley}},\ }\href
  {\doibase 10.1017/S0022112006002631} {\bibfield  {journal} {\bibinfo
  {journal} {J. Fluid Mech.}\ }\textbf {\bibinfo {volume} {568}},\ \bibinfo
  {pages} {119} (\bibinfo {year} {2006})}\BibitemShut {NoStop}%
\bibitem [{\citenamefont {Kuramoto}(1984)}]{Kuramoto_Chemical_1984}%
  \BibitemOpen
  \bibfield  {author} {\bibinfo {author} {\bibfnamefont {Y.}~\bibnamefont
  {Kuramoto}},\ }\href@noop {} {\emph {\bibinfo {title} {Chemical oscillations,
  waves, and turbulence}}}\ (\bibinfo  {publisher} {{Springer-Verlag}},\
  \bibinfo {year} {1984})\BibitemShut {NoStop}%
\bibitem [{Note1()}]{Note1}%
  \BibitemOpen
  \bibinfo {note} {To avoid a collision between a cilium and the basal wall
  when $\alpha _i + \beta _i$ equals $\pi /2$, the root of the cilium is raised
  from the wall by the cylindrical radius as $z_i = r_i/2$.}\BibitemShut
  {Stop}%
\bibitem [{\citenamefont {Pozrikidis}(1992)}]{Pozrikidis_Boundary_1992}%
  \BibitemOpen
  \bibfield  {author} {\bibinfo {author} {\bibfnamefont {C.}~\bibnamefont
  {Pozrikidis}},\ }\href@noop {} {\emph {\bibinfo {title} {Boundary Integral
  and Singularity Methods for Linearized Viscous Flow}}}\ (\bibinfo
  {publisher} {Cambridge University Press},\ \bibinfo {year}
  {1992})\BibitemShut {NoStop}%
\bibitem [{\citenamefont {Blake}(1971)}]{Blake_A_1971}%
  \BibitemOpen
  \bibfield  {author} {\bibinfo {author} {\bibfnamefont {J.}~\bibnamefont
  {Blake}},\ }\href {\doibase 10.1017/S0305004100049902} {\bibfield  {journal}
  {\bibinfo  {journal} {Math. Proc. Cambridge}\ }\textbf {\bibinfo {volume}
  {70}},\ \bibinfo {pages} {303} (\bibinfo {year} {1971})}\BibitemShut
  {NoStop}%
\bibitem [{\citenamefont {Pozrikidis}(1995)}]{Pozrikidis_Finite_1995}%
  \BibitemOpen
  \bibfield  {author} {\bibinfo {author} {\bibfnamefont {C.}~\bibnamefont
  {Pozrikidis}},\ }\href {\doibase 10.1017/S002211209500303X} {\bibfield
  {journal} {\bibinfo  {journal} {J. Fluid Mech.}\ }\textbf {\bibinfo {volume}
  {297}},\ \bibinfo {pages} {123} (\bibinfo {year} {1995})}\BibitemShut
  {NoStop}%
\bibitem [{\citenamefont {Shiratori}\ and\ \citenamefont
  {Hamada}(2006)}]{Shiratori_The_2006}%
  \BibitemOpen
  \bibfield  {author} {\bibinfo {author} {\bibfnamefont {H.}~\bibnamefont
  {Shiratori}}\ and\ \bibinfo {author} {\bibfnamefont {H.}~\bibnamefont
  {Hamada}},\ }\href {\doibase 10.1242/dev.02384} {\bibfield  {journal}
  {\bibinfo  {journal} {Development}\ }\textbf {\bibinfo {volume} {133}},\
  \bibinfo {pages} {2095} (\bibinfo {year} {2006})}\BibitemShut {NoStop}%
\bibitem [{Note2()}]{Note2}%
  \BibitemOpen
  \bibinfo {note} {It cannot be identified whether these are positive or
  negative owing to numerical error.}\BibitemShut {Stop}%
\bibitem [{Note3()}]{Note3}%
  \BibitemOpen
  \bibinfo {note} {Note that because the effect of transformation from $\theta
  _i$ to $\psi _i$ is slight, a qualitative understanding is possible by using
  the original phases $\theta _{1,2}$ for simplicity.}\BibitemShut {Stop}%
\bibitem [{Note4()}]{Note4}%
  \BibitemOpen
  \bibinfo {note} {In addition, $K'_{12}(\theta _1,\theta _2) \approx
  K'_{21}(\theta _1,\theta _2)$ also holds. Precisely, $K'_{12}(\theta
  _1,\theta _2) \not =K'_{21}(\theta _1,\theta _2)$, since the drag torque from
  cilium 1 to 2 does not equal that from cilium 2 to 1 under the model
  considering the cilium length.}\BibitemShut {Stop}%
\end{thebibliography}
%

\end{document}